\documentclass[aps,prd,nofootinbib,12pt]{revtex4-1}
\pdfoutput=1

\makeatletter
\renewcommand{\p@subsection}{}
\renewcommand{\p@subsubsection}{}
\makeatother

\usepackage{graphicx}
\usepackage{amsmath}
\usepackage{slashed}
\usepackage{amssymb}
\usepackage[mathscr]{euscript}
\usepackage{enumerate}
\usepackage{enumitem}
\usepackage{hyperref}
\usepackage{mathtools}

\hypersetup{
	colorlinks=true,
	linktoc=page,
	linkcolor=blue,
	citecolor=blue,
	urlcolor=blue
	}
	
\begin{document}

\title{A holographic proof of the averaged null energy condition}

\author{William R. Kelly}
\email{wkelly@physics.ucsb.edu}
\author{Aron C. Wall}
\email{aroncwall@gmail.com}
\affiliation{University of California at Santa Barbara, Santa Barbara, CA 93106, USA}

\begin{abstract}
The averaged null energy conditions (ANEC) states that, along a complete null curve, the negative energy fluctuations of a quantum field must be balanced by positive energy fluctuations.  We use the AdS/CFT correspondence to prove the ANEC for a class of strongly coupled conformal field theories in flat spacetime.  A violation of the ANEC in the field theory would lead to acausal propagation of signals in the bulk.
\end{abstract}

\maketitle




It has long been known~\cite{Epstein:1965zza} that local quantum field theories allow negative energy fluctuations.  The presence of negative energy is somewhat constrained in theories with a positive total energy; however positivity does not place any obvious restriction on the integrated local energy measured by a single causal observer, and therefore is insufficient to answer many interesting questions.  Among these are the possible existence of warp drives, traversable wormholes, and other exotic phenomena (see e.g.~\cite{Morris:1988tu,Ori:1993eh,Ford:1995wg,Hochberg:1998ii,Visser:2003yf,Alcubierre:1994tu,*Natario:2001tk}) as well as the fate of the singularity theorems of Hawking and Penrose~\cite{Penrose:1964wq,*Hawking:1966sx,*Hawking:1969sw}.

To gain traction on these questions it is necessary to study operators that are better suited to capture the experience of physical observers.  One such operator is the averaged null energy, which is defined as the integral of the null-null component of the stress tensor along a null geodesic which is complete in both directions.  The positivity of this quantity is called the averaged null energy condition (ANEC):
\begin{align} \label{eq:ANEC}
\int_{\gamma(\lambda)} d\lambda \, T_{kk}  \ge 0.
\end{align}
Here $\gamma(\lambda)$ is a complete null geodesic with affine parameter $\lambda$ and associated tangent vector $k$, $T_{ab}$ is the stress tensor, and $T_{kk} := \left< T_{ab} \right> k^a k^b$.

The ANEC was first studied in a purely classical setting by Borde~\cite{Borde:1987qr}, who showed that standard focusing theorems (see~\cite{hawking1975large}) continue to hold when pointwise energy conditions (such as the null energy condition $T_{kk}\ge 0$) are replaced by integrated energy conditions similar to~\eqref{eq:ANEC}.\footnote{See also the earlier work of Tipler~\cite{Tipler:1978zz} on the averaged strong and weak energy conditions.}  Borde's theorems are sufficiently powerful to prove many other results in general relativity including a positive energy theorem~\cite{Penrose:1993ud}, topological censorship~\cite{Friedman:1993ty}, and the Gao-Wald theorem~\cite{Gao:2000ga} (which we review below).  Progress has also been made in proving singularity theorems with weakened energy conditions~\cite{Roman:1986tp,Roman:1988vv,Fewster:2010gm}, though this program remains unfinished.  Some recent reviews of energy conditions are~\cite{Fewster:2012yh,Curiel:2014zba}).

The above results establish that the ANEC is a useful restriction to place on the stress tensor.  It remains to be seen if the ANEC holds for physically interesting field theories.  Existing results establish that the ANEC holds in Minkowski space for free scalar fields~\cite{Klinkhammer:1991ki, Ford:1994bj}, Maxwell fields~\cite{Folacci:1992xg}, and arbitrary two dimensional theories with positive energy and a mass gap \cite{Verch:1999nt}.  One can also use a null surface initial data formulation to show that all free or superrenormalizable theories obey the ANEC in Minkowski space, or on bifurcate Killing horizons~\cite{Wall:2011hj}.

For two dimensional curved spacetimes, one can also prove the ANEC for minimally coupled scalar fields~\cite{Yurtsever:1990gx,Wald:1991xn,Yurtsever:1994wc}, at least if space is noncompact.  Otherwise there is a Casimir energy which allows for ANEC violation in the vacuum, but there is still an ANEC-like bound for energy differences~\cite{Ford:1994bj}.  Many other investigations have provided additional support for the ANEC~\cite{Fewster:2002ne,Graham:2005cq,Fewster:2006uf,Bostelmann:2008hq}, including the work of Blanco and Casini~\cite{Blanco:2013lea} which gives a simple argument showing that negative energy cannot be isolated far away from positive energy in a CFT.

For curved spacetimes with dimension greater than two it is known that the ANEC does not hold on every null curve~\cite{Visser:1996iv,Urban:2010vr,*Urban:2009yt}.  However, Graham and Olum have proposed a weaker condition which they call the `self-consistent achronal ANEC'~\cite{Graham:2007va} (see also~\cite{Flanagan:1996gw}) which weakens~\eqref{eq:ANEC} in two ways.  First,~\eqref{eq:ANEC} is only required to hold only on complete achronal geodesics, i.e. on null curves for which no two points are timelike separated.  Such curves are often called `null lines' in the literature.  Second, the ANEC is only imposed on self-consistent spacetimes for which the gravitational field is sourced by the quantum fields, as well as any additional classical background sources.\footnote{Without this latter restriction there are known violations of the `achronal ANEC'~\cite{Urban:2010vr,*Urban:2009yt}.}  As pointed out in~\cite{Graham:2007va}, generic spacetimes satisfying the self-consistent achronal ANEC will not have any achronal null lines.  But this fact, far from rendering the achronal ANEC trivial, has profound consequences, ruling out closed timelike curves and traversable wormholes~\cite{Graham:2007va, Minguzzi:2008mn}, and also negative energy objects \cite{Penrose:1993ud}.   

But is the self-consistent achronal ANEC true?  So far, Kontou and Olum have also shown that the self-consistent achronal ANEC is satisfied for a minimally coupled free scalar field on a class of curved spacetimes~\cite{Kontou:2012ve}.  At first order in quantum corrections, it also follows if the generalized second law holds on all causal horizons~\cite{Wall:2009wi}.  

In this paper we use the AdS/CFT correspondence~\cite{Maldacena:1997re,Gubser:1998bc,*Witten:1998qj} to prove the ANEC for strongly coupled conformal field theories in $d\ge 2$ spacetime dimensions with a consistent holographic dual.\footnote{For $d=2$ the ANEC follows from an even more general argument.  In 1+1 CFT's the right and left moving sectors decouple and scale invariance implies that the total energy is positive if and only if the left and right Hamiltonians are separately positive---which is equivalent to the ANEC.}  We will consider source-free CFT's in Minkowski space---where all null curves are achronal, and it is neither necessary nor possible to impose gravitational self-consistency.  In a companion paper~\cite{KellyandWall2014} we will extend our results to curved spacetimes.

Unfortunately, it is not currently possible to enumerate all field theories which satisfy the condition of having a consistent holographic dual.  What is known is that AdS/CFT requires a strongly coupled field theory with a large number of species $N$, and several examples of the dual field theories have been worked out in great detail, most famously ${\cal N}=4$ supersymmetric Yang-Mills in four spacetime dimensions.  It has also been conjectured that any strongly coupled CFT with a large-$N$ expansion and a gap in the spectrum of anomalous dimensions has an AdS dual with local dynamics~\cite{Heemskerk:2009pn}.  We will work in the large $N$, strong coupling limit in which the dual theory is well approximated by general relativity.  Note that this limit is distinct from taking the classical limit of the field theory.

The overall strategy of our proof is to assume our theory has nice causal properties and use these properties to derive constraints on the stress tensor.  Our approach is similar in spirit to that of Page et al.~\cite{Page:2002xn}, who proved a positive mass theorem for asymptotically AdS spacetimes with consistent holographic duals.  Their proof is similar to the proofs found in~\cite{Penrose:1993ud,Woolgar:1994ar} except that Page et al. assume their holographic theory has nice causal properties instead of assuming that the bulk spacetime satisfies an energy condition.

Several other researchers have also studied the interplay between bulk causality and various CFT bounds~\cite{Brigante:2007nu,*Brigante:2008gz,Hofman:2008ar,Hofman:2009ug,deBoer:2009pn,*deBoer:2009gx,Camanho:2009vw,*Camanho:2009hu,Camanho:2010ru,Camanho:2014apa}. In~\cite{Brigante:2007nu,*Brigante:2008gz}, Brigante et al. studied the famous viscosity to entropy density ratio $\eta/s$ for conformal fluids with a Gauss-Bonnet gravity dual.  They were able to use causality constraints to place bounds on both the strength of the Gauss-Bonnet coupling and on the ratio $\eta/s$.  These techniques were later generalized and applied to more general Lovelock theories by Camanho et al. in~\cite{Camanho:2010ru}.

In~\cite{Hofman:2008ar}, Hofman and Maldecena derived upper and lower bounds on the ratio of the central charges $a/c$ in a four dimensional CFT.  These bounds are shown to follow from positivity of the energy radiated by collider experiments as measured by distant observers~\cite{Hofman:2008ar} (which is equivalent to the ANEC~\cite{Hofman:2009ug}).  Assuming that the dual bulk is described by an Einstein--Gauss--Bonnet gravity theory, the same lower bound on $a/c$ follows~\cite{Hofman:2009ug} from the assumption that the dual gravitational Lovelock theories satisfies the causality constraint found in \cite{Brigante:2007nu,*Brigante:2008gz}.  This analysis was extended to Lovelock gravity by the authors of~\cite{deBoer:2009pn,*deBoer:2009gx,Camanho:2009vw,*Camanho:2009hu} who also found precise matching between positive energy flux in the boundary and good causal properties in the bulk.  Additionally, Hofman~\cite{Hofman:2009ug} gave a nonrigorous argument that the ANEC should hold in any UV-complete QFT, but this was subject to some unproven assumptions about nonlocal operators in the theory.  Even if there did exist a totally satisfactory field-theoretic proof of the ANEC, it would still be a nontrivial test of AdS/CFT to prove the same result using the duality.

We assume that our theory has good causality properties, in order to prove the ANEC.  This gives a partial converse to~\cite{Hofman:2008ar}, which assumed the ANEC in order to prove that $a/c$ lies in the coupling window that permits good causality.  In the Einstein gravity limit (which in $d = 4$ implies $a/c = 1$), our assumption of good causality is the Gao-Wald theorem, reviewed below.

It is natural to assume the gravity theory is Einstein in light of the recent result of Camanho et al.~\cite{Camanho:2014apa}, who used causality to place a much tighter bound on higher derivative corrections to the bulk equations of motion.  They argue that any finite deviation from Einstein gravity in the bulk at level of the three-point functions (which in $d=4$ is equivalent to a deviation from $a/c=1$) is inconsistent with boundary causality unless the theory contains an infinite tower of massive higher spin particles (as in string theory).  For this reason we will work in the large $N$, strong coupling limit in which these corrections can be neglected.  It would be of interest to extend our analysis to leading order in these corrections.



We now briefly review the elements of the AdS/CFT correspondence that will be used in our proof.  Consider a $d$-dimensional conformal field theory (hereafter called the ``boundary theory") living on Minkowski space, with metric $\eta_{ab}$.  The AdS/CFT correspondence states that this theory has a dual description in terms of a $d+1$ dimensional gravitational theory (the ``bulk" theory) with a metric of the form
\begin{align}
ds^2 = \frac{R_\text{AdS}^2}{z^2} \left( dz^2 + g_{ab}(z,x) dx^a dx^b \right),
\end{align}
where $R_\text{AdS}$ is the AdS length scale and $g_{ab}(0,x) = \eta_{ab}$.  Close to the conformal boundary $z=0$, the Einstein equation dictates that $g_{ab}$ take the form
\begin{align} \label{eq:FGexp}
g_{ab}(z,x) = \eta_{ab} + z^d \gamma_{ab}(z,x)  \, , \qquad \gamma_{ab}(z,x) = t_{ab}(x) + z^2 s_{ab}(z,x) \,
\end{align}
where $t_{ab}$ is a traceless, conserved tensor that is otherwise unconstrained by the equations of motion and $s_{ab}$ is regular at $z=0$.  The AdS/CFT dictionary~\cite{Gubser:1998bc,*Witten:1998qj} states that the expectation value of the stress tensor of the boundary theory is given by
\begin{align} \label{eq:StressTensor}
\left< T_{ab} \right> = \frac{d R_\text{AdS}^{d-1} }{16\pi G } t_{ab},
\end{align}
where $G$ is the $d+1$ dimensional Newton's constant.  From here on we set $R_\text{AdS}=1$; powers of $R_\text{AdS}$ can be restored by dimensional analysis.  In writing down~\eqref{eq:FGexp} and~\eqref{eq:StressTensor} we have used our restriction that all boundary sources have been turned off.  In the bulk, this amounts to requiring that any bulk matter fields fall of fast enough at conformal infinity that they do not play a direct role in our analysis.

\begin{figure}
\includegraphics[width=0.4 \textwidth]{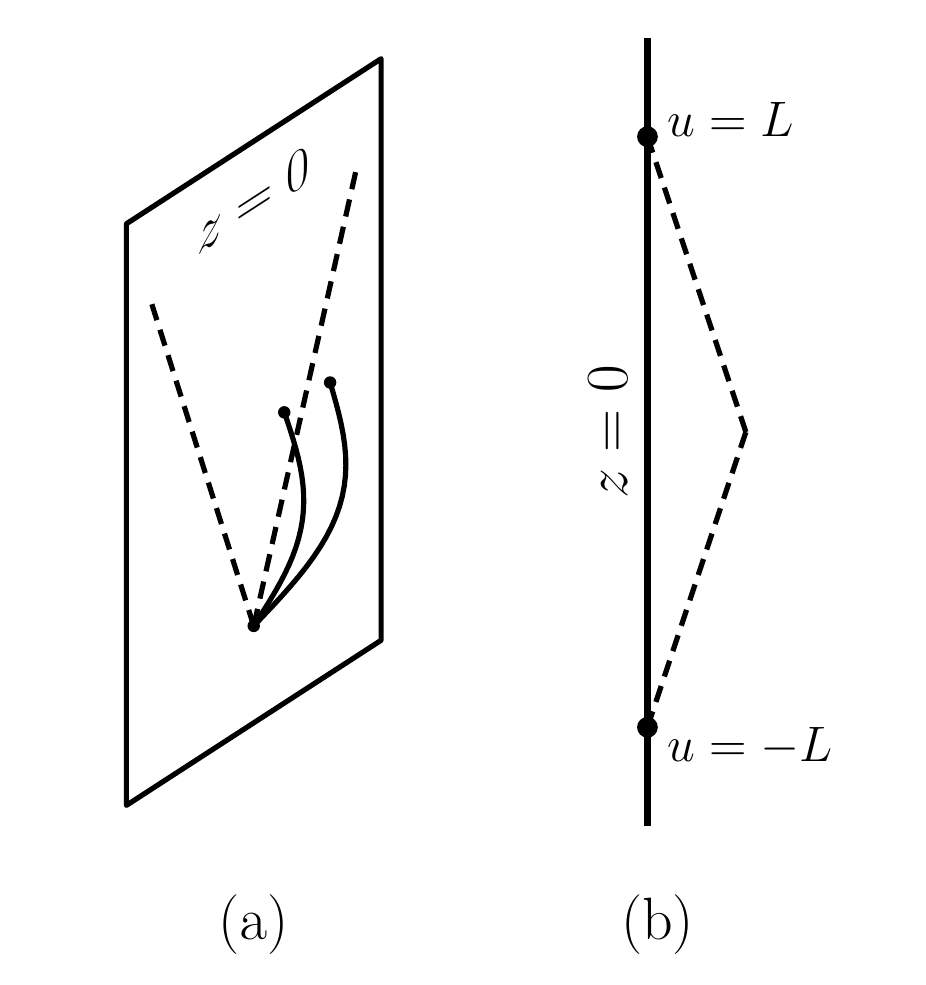} 
\caption{(a) Two curves which begin and end on the conformal boundary but which dip into the bulk.  The assumption of good causality requires that the curve which ends outside of the boundary light cone (dashed line) cannot be causal. (b) Schematic of the construction used in our proof.  The solid line is the conformal boundary $z=0$ and the dashed lines represent causal curves extending into the bulk.  The $v$ direction has been suppressed in this diagram.}
\label{fig:geodesic}
\end{figure}

In order for the boundary theory to be local there can be no ``shortcuts through the bulk" which would effectively allow signals to propagate faster than light (see Fig.~\ref{fig:geodesic}(a)).  This principle is encapsulated by the Gao-Wald theorem (Theorem 2 of~\cite{Gao:2000ga}), which states that the fastest possible path between two boundary points is a null geodesic on the boundary.  The Gao-Wald theorem was proven for Einstein gravity whenever the \emph{bulk} stress tensor $T^\text{bulk}_{\mu\nu}$ satisfies the ANEC and  the bulk is a generic, asymptotically locally AdS spacetime.  For our purposes it is natural to take the \emph{conclusion} of the Gao-Wald theorem to be part of the definition of a consistent holographic theory.  After all, if the bulk dual permitted signaling through the bulk faster than the speed of light on the boundary, it would imply that the dual CFT permits acausal signaling (see e.g.~\cite{Horowitz:1999gf}).  Alternatively, we could assume that our classical bulk geometry satisfies the assumptions of the Gao-Wald and invoke the theorem.

Finally our proof requires two formal assumptions about $T_{kk}$, namely that $|T_{kk}|$ is bounded ($|T_{kk}| < T_\text{max}$) and that $T_{kk}$ and its derivatives are absolutely convergent on $\gamma(\lambda)$ (i.e. that $\int_\gamma |T_{kk}|$, $\int_\gamma |\partial T_{kk}|$, $\int_\gamma |\partial^2 T_{kk}| ,\dots $ are finite).  This allows us to define the integral~\eqref{eq:ANEC} as a limit of integrals over finite intervals.  It is likely that these assumptions could be weakened by using the more general formulation of the ANEC in e.g.~\cite{Borde:1987qr,Wald:1991xn}.


We are now ready to begin our proof.  Consider null coordinates on the boundary spacetime
\begin{align}
\eta_{ab}\,dx^a\,dx^b = - (du\,dv + dv\,du) + d\vec{y}^{\phantom{,}2}
\end{align}
where $d\vec{y}^{\phantom{,}2}$ is the Euclidean line element over the remaining $d-2$ spatial directions.  Note that $u$ is an affine parameter for the geodesic $v = (\text{constant}),\, \vec y = (\text{constant})$.  We assume that all components of the bulk metric are smooth and bounded in these coordinates.

The strategy of our proof is to construct a causal curve which dips into the bulk, but has both endpoints anchored to the boundary.  We will engineer this curve to remain close to the boundary and calculate the time delay or advance relative to a nearby boundary null geodesic (see Fig.~\ref{fig:geodesic}(b)).  We will find a positive ``kinetic'' contribution to the time delay coming from the radial motion of the curve into the bulk, and a second ``potential'' contribution whose sign is that of $t_{uu}$, and therefore may be either a delay or advance.  We will carefully construct our curve so that the latter contribution dominates.  Our causal assumption requires that the net time delay of the entire excursion must be positive; we will show that this restriction implies~\eqref{eq:ANEC}.

We parameterize our curve by the coordinate $u$ so that $v = V(u)$ and $z=Z(u)$.  Without loss of generality we set $\vec y = 0$.  This curve will be causal if $V,Z$ satisfy
\begin{align} \label{eq:NullCondition}
(Z')^2 -2 V' + Z^d \left( \gamma_{uu} +\gamma_{uv} \, V' + \gamma_{vv} \, (V')^2 \right)   \le 0,
\end{align}
where primes indicate $u$-derivatives.  

We now construct a curve satisfying~\eqref{eq:NullCondition}.  Consider the interval $ u \in [-L,L]$ for some $L$ which we will ultimately take to be arbitrarily large.  It is useful to introduce a small parameter $\epsilon$, which parameterizes how deep into the bulk our curve reaches.  We need to take an $\epsilon \to 0$ limit in order to relate our results to $t_{uu}$ using~\eqref{eq:FGexp}, but in this limit any time advance due to $t_{uu}$ is swamped by the time-delay due to veering into the bulk.  Thus in order to prove an interesting result it is necessary to take a simultaneous limit in which $L$ becomes large as $\epsilon$ becomes small.  This is why good causality implies the ANEC but not the null energy condition $T_{uu} \ge 0$.  It turns out to be convenient to set
\begin{align} \label{eq:epsDef}
L = \epsilon^{-(d-2+2\alpha)}\, ,
\end{align}
where $\alpha$ is a constant satisfying $0< \alpha<2/3$.  We will construct our casual curve by joining together two smooth causal curves at a sharp angle, one curve dipping into the bulk and the other coming back to the boundary (Fig.~\ref{fig:geodesic}(b)), by choosing $V,Z$ to be given on the interval $u\in[-L,L]$ by
\begin{align} \label{eq:CausalCurve}
Z(u) &=\epsilon \left(\frac{L-|u|}{L}\right) \cr
 V(u) &= \frac{1}{2 }\left[ \frac{\epsilon^{2-\alpha}}{L} \left(\frac{L+u}{L}\right)+ \epsilon^d \int_{-L}^u du'  \left(\frac{L-|u'|}{L}\right)^d \gamma_{uu}(u',0) \right] \, ,
\end{align} 
(In the second equation, the first term is the ``kinetic'' time delay and the second the ``potential''  delay.)  The appearance of $\alpha$ in the exponent of the first term represents an extra time delay we have inserted to ensure that~\eqref{eq:NullCondition} is satisfied for sufficiently small $\epsilon$ (keeping $\alpha$ fixed).  We have used the fact that $\gamma_{uu}$ is smooth to power expand:
\begin{align}
\gamma_{uu}(u, V(u)) = \gamma_{uu}(u,0) + O(\epsilon^d),
\end{align}
since $V(u)\sim \epsilon^d$.\footnote{$V(u)\sim \epsilon^d$ because the integral in~\eqref{eq:CausalCurve} remains finite as $L\to \infty$.  This follows from the arguments given below~\eqref{eq:AlmostANEC}.}

Since the curve~\eqref{eq:CausalCurve} is causal, our causality assumption requires that the end points of~\eqref{eq:CausalCurve} must be causally separated in the boundary spacetime.  This implies that the time delay $\Delta V := V(L) - V(-L)$ must be positive.  In terms of the stress tensor~\eqref{eq:StressTensor} we then find that for any $L$
\begin{align} \label{eq:AlmostANEC}
\int_{-L}^L d\lambda \, f_L \, T_{kk}  \ge  -  \left(\frac{16\pi G}{d R_\text{AdS}^{d-1} }\right)   \left( 2\epsilon^\alpha +  \int_{-L}^L d\lambda \,  \epsilon^2 |s_{kk}| \right)\,, \quad f_L(\lambda) = \left( \frac{L- \left|\lambda\right|}{L}\right)^d \, ,
\end{align} 
where we have momentarily restored the correct powers of $R_\text{AdS}$.  Note that $0 \le f_L \le 1$.  We will now show that~\eqref{eq:AlmostANEC} implies the ANEC~\eqref{eq:ANEC}.

First, we argue that  $ \int_{-L}^L d\lambda \, \epsilon^2 |s_{kk}|$ vanishes in the limit $L\to \infty$.  Expanding the Einstein equation about $z=0$ allows us to write $s_{kk}$ as an algebraic (non-linear) function of $t_{ab}$ and its derivatives.\footnote{See, for example, Eq.~(7) in~\cite{Henningson:1998gx}.}  The contribution to the integrand from quadratic and higher order terms vanish like $\epsilon^d$ by power counting.   Because we assume the metric components are bounded, the contribution to the integral from these terms must scale like $\epsilon^d L  = \epsilon^{2(1-\alpha)}$ which vanishes as we take $L\to \infty$.  The terms in $s_{kk}$ that are linear in $t_{kk}$ have finite integrals by our assumption that $T_{kk}$ and its derivatives are all absolutely convergent, therefore the contribution from these terms vanishes like $\epsilon^2$.  Finally, terms proportional to $\eta^{ab} t_{ab}$ vanish because $t_{ab}$ is traceless.  This accounts for all possible contributions to $s_{kk}$, therefore the right hand side of~\eqref{eq:AlmostANEC} vanishes as $L\to\infty$.

For illustrative purposes we now treat the simple case where $T_{kk}$ is non-negative outside of some interval $\lambda \in [-\lambda_0 , \lambda_0]$.  In this case we may write
\begin{align} \label{eq:TmaxIntegral}
\int_{-L}^L d\lambda \, f_L \, T_{kk}  &\le -  T_\text{min}^{(\lambda_0)} \left[ \int_{-\lambda_0}^{\lambda_0} d\lambda (1-f_L)\right] + \int_{-L}^L d\lambda  \, T_{kk}  ,
\end{align}
where $T_\text{min}^{(\lambda_0)}$ is a lower bound on $T_{kk}$ in $ [-\lambda_0 , \lambda_0]$, which must exist by our assumption that $|T_{kk}|$ is bounded.  For fixed $\lambda_0$ the term in square brackets vanishes like $L^{-2}$ as $L$ becomes large.  Combining~\eqref{eq:TmaxIntegral} and~\eqref{eq:AlmostANEC} and taking $L\rightarrow\infty$ yields~\eqref{eq:ANEC}.

If the previous assumption doesn't hold then the integral in~\eqref{eq:ANEC} is oscillatory and we must be a little more careful.  In this case it is useful to note that
\begin{align}  \label{eq:almostANEC2}
\int_{-L}^L d\lambda \, f_L \,T_{kk} 
\le \int_{-L}^L d\lambda (1-f_L) |T_{kk}| + \int_{-L}^L d\lambda \, T_{kk}.
\end{align}
We now must show that the first term on the right hand side of~\eqref{eq:almostANEC2} vanishes as $L\rightarrow\infty$ and~\eqref{eq:ANEC} will follow as before.  In other words, we must show that for any $\delta>0$ there exists an $L$ such that
\begin{align} \label{eq:deltaLimit}
 \int_{-L}^L d\lambda (1-f_L) |T_{kk}| < \delta.
\end{align}
By our assumption that $T_{kk}$ is absolutely convergent, there must exist some $\lambda_1$ such that
\begin{align}
\int_{\lambda_1}^\infty d \lambda \, |T_{kk}| + \int_{-\infty}^{-\lambda_1} d \lambda \, |T_{kk}| < \frac{\delta}{2}.
\end{align}
Now for any $L>\lambda_1$ we have
\begin{align}
 \int_{-L}^L d\lambda (1-f_L) |T_{kk}|
< T_\text{max}^{(\lambda_1)}  \left[\int_{-\lambda_1}^{\lambda_1} d\lambda(1-f_L) \right] + \frac{\delta}{2}  ,
\end{align}
where $T_\text{max}^{(\lambda_1)}$ is the maximum of $|T_{kk}|$ in $[-\lambda_1,\lambda_1]$.  As before the term in square brackets goes like $L^{-2}$, and therefore there always exists some $L$ satisfying~\eqref{eq:deltaLimit}.  This completes our proof of~\eqref{eq:ANEC}.

We have just given a simple, geometric proof of the ANEC for any field theory on Minkowski space with a consistent holographic dual.  Our proof applies to strongly coupled CFT's on Minkowski space, but it would be of interest to extend our results to curved space as a test of the self-consistent achronal ANEC~\cite{Graham:2007va}.  On a curved background Eqs.~\eqref{eq:FGexp} and~\eqref{eq:StressTensor} contain extra terms that involve the background metric and curvature as well as any background source terms.  These terms become increasingly complicated as the dimension increases and there is no known expression for arbitrary dimension.  However, all of the curvature terms needed to analyze $d\le 6$ have been known for some time (see~\cite{deHaro:2000xn})---six dimensions being the largest dimension with a known AdS/CFT duality~\cite{Maldacena:1997re}.  For general backgrounds the analysis becomes more complex and we defer the details to~\cite{KellyandWall2014}.

It would also be of interest to extend our arguments to include perturbative quantum and stringy corrections in the bulk.  Because we are proving an inequality we only need to consider perturbative corrections when the classical inequality is saturated.  Presumably the ANEC can only be saturated in very stringent situations, but this does not follow from our proof.  It may be possible to make progress on this point by bounding the minimum time delay for a generic spacetime, possibly using techniques adapted from~\cite{Gao:2000ga,Page:2002xn}.

These results have the potential to lead to new insights about holography in the spirit of~\cite{Page:2002xn,Woolgar:1994ar,Brigante:2007nu,*Brigante:2008gz,Hofman:2008ar,Hofman:2009ug,deBoer:2009pn,*deBoer:2009gx,Camanho:2009vw,*Camanho:2009hu,Camanho:2010ru,Camanho:2014apa}.  There are many unanswered questions about the emergence of causal structure in AdS/CFT,  so understanding the field-theoretic origin of the Gao-Wald theorem---and any perturbative higher-curvature analogues---will lead to new insights related to this emergence.  It would be of interest to develop a more complete understanding of how bulk causality restricts the field theory.  Our analysis was restricted to causal curves which remain close to the boundary, but curves which go deeper into the bulk place restrictions on the fields in bounded regions, which are nonlinear in the boundary stress-tensor.  

\section*{Acknowledgements}

It is a pleasure to thank Netta Engelhardt, Gary Horowitz, Don Marolf, and Ken Olum for helpful discussions.  We also thank Hongbao Zhang for identifying an error in a previous version of this paper.  This work was supported in part by the National Science Foundation under Grant No PHY11-25915, by FQXi grant FRP3-1338, and by funds from the University of California.  AW was also supported by the Simons Foundation.

\bibliographystyle{kp}

\bibliography{holographicANECv8.bbl}

\end{document}